\documentclass[times,twocolumn,final]{elsarticle}

\usepackage{medima}
\usepackage{framed,multirow}

\usepackage{amssymb}
\usepackage{latexsym}

\usepackage{url}
\usepackage{xcolor}
\usepackage[hidelinks]{hyperref}
\definecolor{newcolor}{rgb}{.8,.349,.1}

\usepackage{mathtools}
\usepackage{amsmath}

\newcommand\Tstrut{\rule{0pt}{2.0ex}}         
\newcommand\Bstrut{\rule[-0.9ex]{0pt}{0pt}}   

\usepackage[ruled,vlined,algo2e]{algorithm2e}
\SetKwComment{Comment}{$\triangleright$\ }{}

\usepackage{hyperref}

\journal{Medical Image Analysis}

\begin{document}

\verso{B. Zhou \textit{et~al.}}

\begin{frontmatter}

\title{Cascaded Multi-path Shortcut Diffusion Model for Medical Image Translation}

\author[1]{Yinchi \snm{Zhou}\corref{cor1}}
\ead{yinchi.zhou@yale.edu}
\author[4]{Tianqi \snm{Chen}}
\author[4]{Jun \snm{Hou}}
\author[1]{Huidong \snm{Xie}}
\author[1,2]{Nicha C. \snm{Dvornek}}
\author[5]{S. Kevin \snm{Zhou}}
\author[6]{David L. \snm{Wilson}}
\author[1,2,3]{James S. \snm{Duncan}}
\author[1,2]{Chi \snm{Liu}}
\author[7]{Bo \snm{Zhou}\corref{cor1}}
\cortext[cor1]{Corresponding author.}
\ead{bo.zhou@northwestern.edu}

\address[1]{Department of Biomedical Engineering, Yale University, New Haven, CT, USA}
\address[2]{Department of Radiology and Biomedical Imaging, Yale School of Medicine, New
Haven, CT, USA}
\address[3]{Department of Electrical Engineering, Yale University, New Haven, CT, USA.}
\address[4]{Department of Computer Science, University of California Irvine, Irvine, CA, USA}
\address[5]{School of Biomedical Engineering \& Suzhou Institute for Advanced Research, University of Science and Technology of China, Suzhou, China}
\address[6]{Department of Biomedical Engineering, Case Western Reserve University, Cleveland, OH, USA.}
\address[7]{Department of Radiology, Northwestern University, Chicago, IL, USA}





\begin{abstract}
Image-to-image translation is a vital component in medical imaging processing, with many uses in a wide range of imaging modalities and clinical scenarios. Previous methods include Generative Adversarial Networks (GANs) and Diffusion Models (DMs), which offer realism but suffer from instability and lack uncertainty estimation. Even though both GAN and DM methods have individually exhibited their capability in medical image translation tasks, the potential of combining a GAN and DM to further improve translation performance and to enable uncertainty estimation remains largely unexplored. In this work, we address these challenges by proposing a Cascade Multi-path Shortcut Diffusion Model (CMDM) for high-quality medical image translation and uncertainty estimation. To reduce the required number of iterations and ensure robust performance, our method first obtains a conditional GAN-generated prior image that will be used for the efficient reverse translation with a DM in the subsequent step. Additionally, a multi-path shortcut diffusion strategy is employed to refine translation results and estimate uncertainty. A cascaded pipeline further enhances translation quality, incorporating residual averaging between cascades. We collected three different medical image datasets with two sub-tasks for each dataset to test the generalizability of our approach. Our experimental results found that CMDM can produce high-quality translations comparable to state-of-the-art methods while providing reasonable uncertainty estimations that correlate well with the translation error.

\end{abstract}


\end{frontmatter}


\section{Introduction}
Image-to-image translation (I2I) plays an important role in medical imaging with wide applications in different medical imaging modalities, such as Digital Radiography (DR), Computed Tomography (CT), and Magnetic Resonance Imaging (MRI). The applications can be summarized into both intra-modality I2I and inter-modality I2I in medical imaging. In the applications of medical X-ray, intra-modality I2I can achieve the high-quality reconstruction of images under radiation dose reduction scenarios. For example, CT radiation dose reduction can be accomplished by translating the sparse-view CT, i.e. acquired with a reduced number of projection views, into the full-view CT \citep{zhou2021limited,zhang2018sparse,wu2021drone}. Dual-energy (DE) DR radiation dose can be reduced by nearly half by translating the standard single-shot DR into two-shot DE images, i.e. soft-tissue and bone images \citep{zhou2019generation,yang2017cascade,liu2023bone}. In MRI applications, intra-modality I2I can be used for image acquisition acceleration. For example, one can use T1 to assist the synthesis/reconstruction of T2 and FLAIR with no or undersampled k-space data \citep{yang2020mri,zhou2020dudornet}. In the application of CT-free PET or SPECT attenuation correction, inter-modality I2I that translates PET or SPECT into CT also helps remove the need for CT acquisition, thus reducing the overall radiation dose \citep{zhou2024pour,chen2022direct,chen2022cross}. Therefore, building an accurate and robust I2I method that is generalizable to a wide range of medical imaging applications is important.

With the recent advancements in deep learning (DL), many DL-based I2I methods have been proposed and adapted to the medical imaging field, demonstrating promising performance. In general, prior I2I methods can be summarized into two classes: Generative Adversarial Network (GAN) and Diffusion Model (DM). 

With paired training data available for I2I, one of the most widely used I2I GANs is the conditional GAN (cGAN \citep{isola2017image}), which consists of 1) a generator that aims to translate an input image into a target image, and 2) a discriminator that conditions on the initial input and the translation for adversarial training. A large amount of cGAN variants have been developed for various medical imaging applications. For example, \cite{huang2021gan} proposed a GAN with dual discriminators on both image and gradient domains for low-dose CT (LDCT) to full-dose CT (FDCT) translation. \cite{denck2021mr} proposed a cGAN with an additional input of MRI acquisition information for intra-MRI-modality translations. \cite{nie2018medical} proposed to modify the cGAN with the addition of a gradient-based loss function, and showed successful applications in MRI to CT translation and 3T-MRI to 7T-MRI translation. Based on this, \cite{zhou2019generation} further designed a multi-scale cGAN for single-shot DR image to DE image translation. In PET, \cite{gong2020parameter} also proposed a GAN with parameter transferring for low-dose PET (LDPET) to full-dose PET (FDPET) translation. Even though reasonable translation performance can be achieved with simple and fast one-step inference from the generator, training GANs can be challenging due to the need to balance between the optimization of the generator and discriminator (e.g. find the saddle point of the min-max objective). The training is therefore susceptible to non-convergence and mode collapse.   
\begin{figure}[htb!]
\centering
\includegraphics[width=0.475\textwidth]{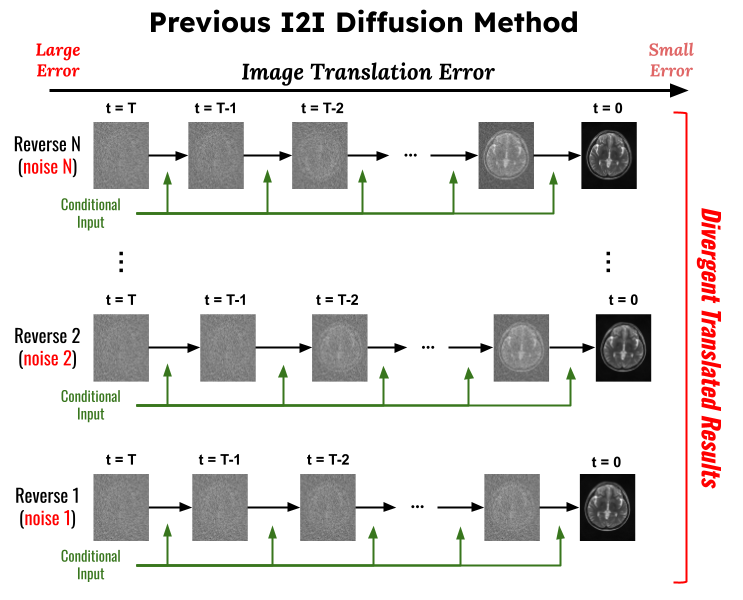}
\caption{\textcolor{black}{Illustration of previous I2I diffusion model generation process. Starting the reverse process with different noise initialization leads to divergent translation results.}}
\label{fig:previous}
\end{figure}

\begin{figure*}[htb!]
\centering
\includegraphics[width=0.92\textwidth]{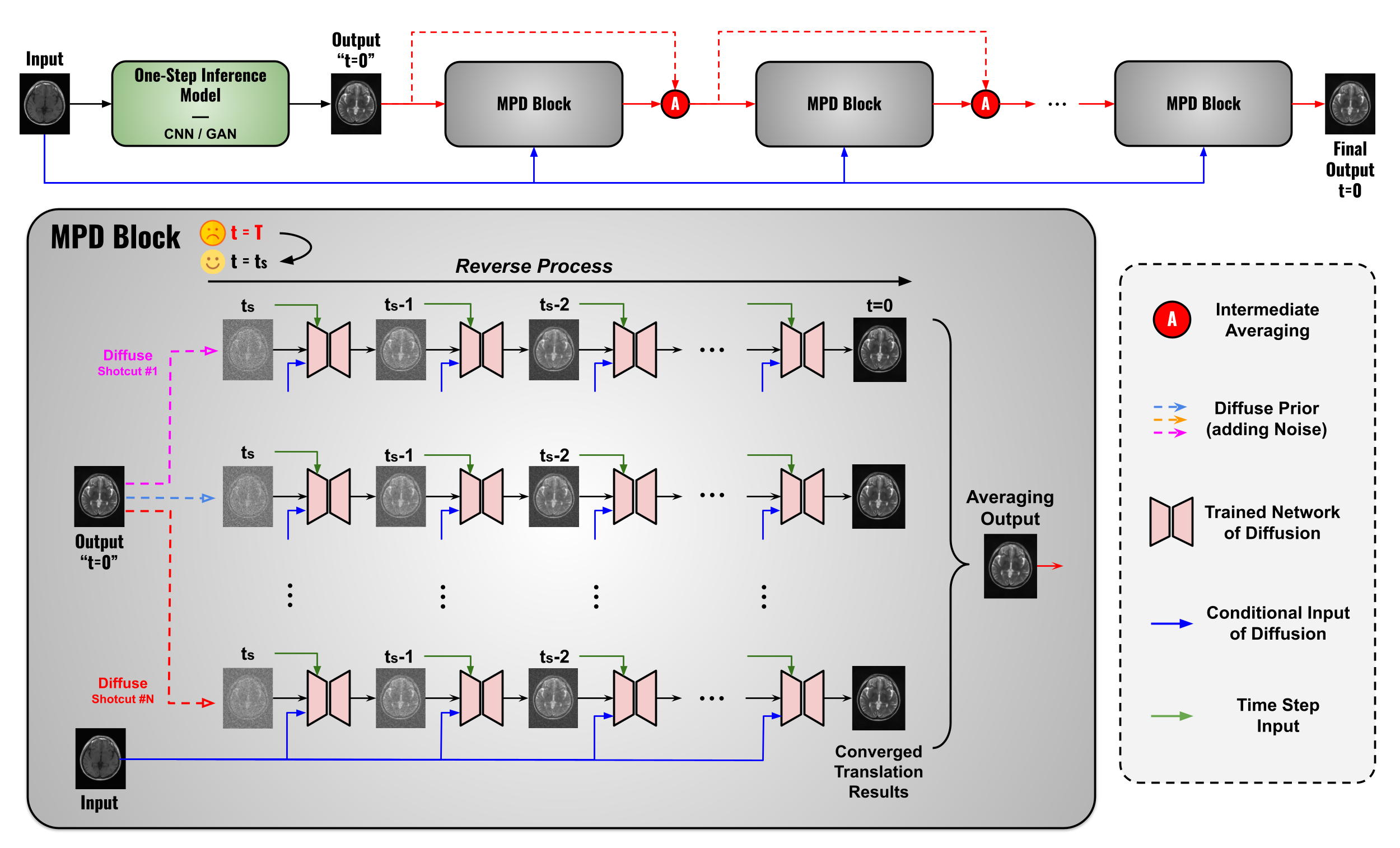}
\caption{The overall workflow of our proposed Cascade Multi-path Shortcut Diffusion Model (CMDM). CMDM consists of a one-step inference model (green) and cascades of MPD block (grey). Each MPD block consists of multiple shortcut reverse paths starting with a prior image with different noise. The cascades are connected with residual averaging operations.}
\label{fig:network}
\end{figure*}

On the other hand, I2I diffusion models have been recently developed and show superior performance than GANs. For general-purpose I2I with DM, \cite{saharia2022palette} proposed a unified framework, Palette, which adds conditional image inputs to the previously developed Denoising Diffusion Probabilistic Model (DDPM \citep{ho2020denoising}), thus enabling the I2I functionality of DDPM. To reduce the randomized initialization process and improve the stability in I2I DM, direct bridging diffusion methods have been investigated. Notably, \cite{li2023bbdm} developed a Brownian Bridge Diffusion Model (BBDM) that learns the translation between two domains directly through the bidirectional diffusion process, i.e. Brownian Bridge, rather than a conditional generation process. Similarly, \cite{liu20232} proposed a Schrodinger Bridge I2I Diffusion Model (I2SB) that directly learns the nonlinear diffusion processes between two domains. Both had shown improved I2I performance in natural image translation tasks. Similar to I2I GANs, these DM methods have been applied in medical imaging. For example, \cite{moghadam2023morphology} utilized DDPM to synthesize artificial histopathology images with rare cancer subtypes to mitigate data imbalance problems for medical data. \cite{lyu2022conversion} proposed to translate CT into MRI with conditional DDPM and score-matching models. The forward and backward diffusion processes are guided by T2 MRI. \cite{gong2023pet} proposed to perform brain PET image denoising with MRI as prior information to improve image quality. \cite{gao2023corediff} utilized a contextual contained network in the DM to improve the LDCT denoising. Furthermore, 2D DMs have also been employed for 3D translation tasks, including low-count PET image denoising \citep{xie2023dose}, CT reconstruction \citep{chung2023solving}, and MRI super-resolution and reconstruction \citep{lee2023improving}. Direct extension to 3D DM were also explored \citep{pan2023synthetic}. However, there are several unique challenges of DM for I2I. First, those methods require iterating over a large number of steps in the reverse process, and most methods start the generation with pure random noise \citep{saharia2022palette,lyu2022conversion,gong2023pet,xie2023dose,chung2023solving,lee2023improving}. This protocol not only significantly slows down the translation speed, but could also lead to diverged and sub-optimal translation results if different random noise initialization were used in the input when running multiple reverse runs (Figure \ref{fig:previous}). Even though direct bridging methods \citep{li2023bbdm,liu20232} are translation deterministic given that no random noise input is used, they still require a large number of reverse iteration steps. Another challenge of the deterministic translation is that they also cannot generate translation uncertainty maps which is crucial for medical images, since the model's prediction error can be used to pinpoint problem areas or give clinicians more information \citep{shi2021inconsistency,jungo2019assessing,wolleb2022diffusion}. It is then a unique advantage of the stochastic sampling process of the conditional DDPM \citep{saharia2022palette} to obtain the uncertainty map through running the DM repeatedly with multiple random noises \citep{wolleb2022diffusion}. Therefore, it is desirable to develop an I2I DM method that can generate high-quality converged translation results with a reduced number of required iterations, while also being able to provide translation uncertainty estimation. 

Looking into prior works, even though both GAN and DM methods have individually exhibited their capability in medical image translation tasks, the potential of combining GAN and DM for further improving translation performance remains largely unexplored. With this and to address the aforementioned challenges in DM, we proposed a Cascade Multi-path Shortcut Diffusion Model (CMDM) for medical image-to-image translation in this work. Specifically, CMDM consists of three key components. Firstly, we proposed to utilize a cGAN-generated prior image with diffusion (i.e. noise addition) for providing an arbitrary time point's input in the reverse process. With this shortcut strategy, 1) we need fewer number of iterations thus reducing the processing time, and 2) the reverse process starts with prior information from cGAN instead of pure noise, thus leading to more consistent and robust performance. Second, we proposed to perform this shortcut reverse process multiple times with different noise additions to the cGAN-generated prior. Then, refined translation can be obtained by averaging the multi-path shortcut diffusion results. Meanwhile, the translation uncertainty can also be estimated by computing the standard deviation of the multi-path shortcut diffusion results. Lastly, to further refine the translation, we devised a cascade pipeline with the multi-path shortcut diffusion embedded in each cascade. Between each cascade, we used a residual averaging strategy where each cascade's prior image is perturbed by averaging the last cascade's output and the previous prior image. We collected three datasets in different medical imaging modalities with different image translation applications. Our experimental results on these datasets demonstrated that we can generate high-quality translation images, competitive with the prior state-of-the-art I2I methods. We also show our method can generate reasonable uncertainty estimation that correlates well with the translation error.

\section{Methods}
\subsection{Cascaded Multi-path Shortcut Diffusion Model}


The overall architecture of the Cascaded Multi-path Shortcut Diffusion Model (CMDM) is illustrated in Figure \ref{fig:network}. The CMDM consists of (1) a one-step inference model, i.e. cGAN \citep{isola2017image}, for generating a prior translation image, and (2) a conditional denoising diffusion probabilistic model (cDDPM) to further refine the prior translation image in a cascade and multi-path fashion. The training and inference details are as follows.

\vspace{0.1cm}
\noindent\textbf{Training:} Let us denote the input image as $x$ and the translation target as $y_0$. For the prior image generation part, we utilized a generative network, i.e. UNet \citep{ronneberger2015u}, that aims to predict $y_0$ from $x$. \textcolor{black}{The network can be trained in a conditional adversarial fashion \citep{isola2017image} using both a pixel-wise L2 loss} 
\begin{equation} 
    \mathcal{L}_{gen} = || f_{prior}(x) - y_0 ||_2^2,
\end{equation}
\textcolor{black}{and a conditional adversarial loss}
\begin{equation}
    \mathcal{L}_{adv} = - log (f_{adv}(y_0 | x)) - log (1 - f_{adv}(f_{prior}(x) | x)), 
\end{equation}
where $f_{prior}(\cdot)$ is the generative network for generating the prior image and $f_{adv}(\cdot)$ is the discriminator network. 

\begin{algorithm2e*}
\caption{Inference Process - Cascaded Multi-path Shortcut Diffusion Model (CMDM)}
\label{alg:test}
\(\textbf{Input: } x\in N^{d_1\times d_2}\)\\
\(\textbf{Initialize \#1: }t_s\in[0, T]\text{: the start timestep of denoising process}\)\\
\(\textbf{Initialize \#2: }N_c\text{: the number of cascades}; N_p\text{: the number of shortcut paths}\) \\
\(\textbf{Initialize \#3: }f_{prior}(\cdot)\text{: prior image generation network}; f_{dm}(\cdot)\text{: conditional diffusion network}\) \\
\For{$c = 1, 2, 3, ..., N_c$}{
    \If{$c = 1$}{
    $y^c_0 = f_{prior}(x)$ \Comment*[r]{Prior image generation by one-step CNN inference}
    }
    \Else{
     $y^c_0 = \frac{1}{2}(y^{avg}_0 + y^{c-1}_0)$ \Comment*[r]{Subsequent prior image generation by residual averaging}
    }
    
    \For{$p = 1, 2, 3, ..., N_p$}{
        $y^p_{t_s} = \sqrt{\gamma_{t_s}} y^c_0 + \sqrt{1-\gamma_{t_s}} \epsilon_p$ \text{,} $\epsilon_p \sim \mathcal{N}(0, I)$ \Comment*[r]{Adding noise to the prior image for shortcut at $t_s$}
        
        \For{$t = t_s, t_s-1, t_s-2, ..., 1$}{
            $\epsilon_t \sim \mathcal{N}(0, I)$ \Comment*[r]{Sampling noise in the reverse process}
 
            $y^p_{t-1} = \frac{1}{\sqrt{\alpha_t}} (y^p_t - \frac{1-\alpha_t}{\sqrt{1-\gamma_t}} f_{dm} (x, y^p_t, \gamma_t)) + \sqrt{1-\alpha_t} \epsilon_t$ \Comment*[r]{Iterative reverse process in a single path}
        }
    }
    $y^{avg}_0 = \frac{1}{N_p} \sum^{N_p}_{p=1} y^p_0$  \Comment*[r]{Averaging the multiple shortcut paths outputs}
}
\Return{$y^{avg}_0$} \Comment*[r]{Outputting the last cascade's multi-path averaging result}
\end{algorithm2e*}

On the other hand, the diffusion model consists of a forward diffusion process and a reverse denoising process. The forward diffusion process is a Markovian process that gradually adds Gaussian noise to the target image $y_0$ over $T$ iterations, and can be defined as:
\begin{equation}
    q(y_{1:T}| y_0) = \prod^T_{t=1} q(y_t | y_{t-1}),
\end{equation}
where $q(y_{t} | y_{t-1}) = \mathcal{N} (y_{t-1} ; \sqrt{\alpha_t} y_{t-1}, (1-\alpha_t) I)$, and $\alpha_t$ are the noise schedule parameters. $T$ is empirically set to $1000$ here such that $y_T$ is visually indistinguishable from Gaussian noise. Then, the forward process can be further marginalized at each step as:
\begin{equation}
    q(y_t | y_0) =  \mathcal{N}(y_t ; \sqrt{\gamma_t} y_0, (1 - \gamma_t) I),
\end{equation}
where $\gamma_t = \prod^t_{s=0} \alpha_s$. Then, the posterior distribution of $y_{t-1}$ given $(y_0,y_t)$ can be derived as:
\begin{equation} \label{eq:posterior}
    q(y_{t-1} | y_0, y_t) = \mathcal{N} (y_{t-1} | \mu, \sigma^2 I),
\end{equation}
where $\mu=\frac{\sqrt{\gamma_{t-1}}(1-\alpha_t)}{1-\gamma_t}y_0 + \frac{\sqrt{\alpha_t}(1-\gamma_{t-1})}{1-\gamma_t}y_t$ and $\sigma^2 = \frac{(1-\gamma_{t-1})(1-\alpha_t)}{1-\gamma_t}$. With this, the noisy image during the forward process can thus be written as
\begin{equation} \label{eq:addnoise}
    \hat{y}_t = \sqrt{\gamma_t}y_0 + \sqrt{1-\gamma_t}\epsilon 
\end{equation}
where $\epsilon \sim \mathcal{N}(0,I)$. Here, the goal is to estimate the noise and thus gradually remove it during the reverse process to recover the target image $y_0$. In our conditional diffusion model, we utilized another generative network $f_{dm}(\cdot)$ to estimate the noise with another pixel-wise L2 loss
\begin{equation}
    \mathcal{L}_{dm} = ||f_{dm} (x,  \hat{y}_t, \gamma_t) - \epsilon||^2_2
\end{equation}
where $x$ is the input image that is also used as conditional input here. $\hat{y}_t$ is the noisy image, and $\gamma_t$ is the current noise level. The prior image generation network and the diffusion model network were trained separately. 

\begin{figure*}[htb!]
\centering
\includegraphics[width=0.92\textwidth]{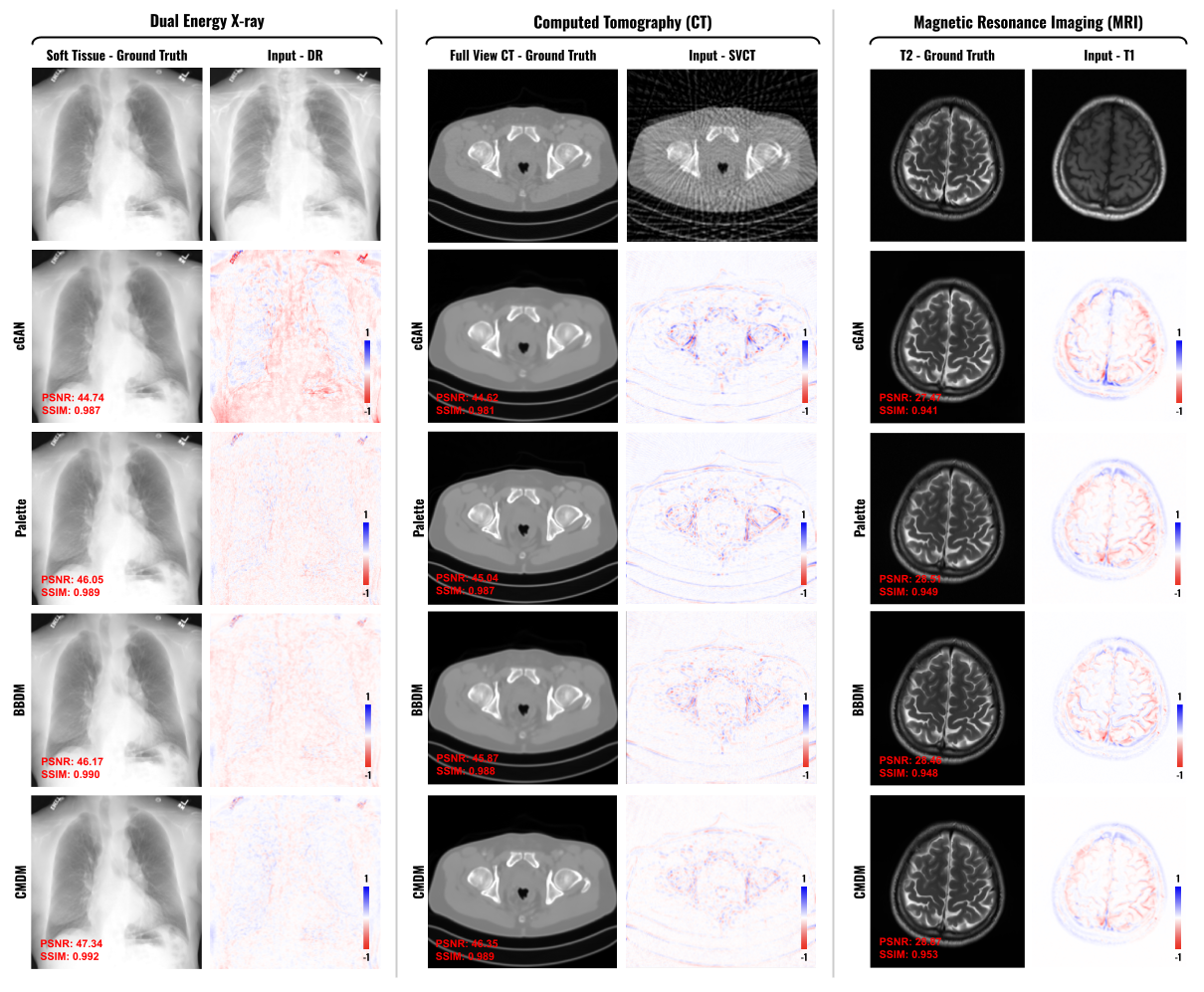}
\caption{Qualitative comparison of translation results and corresponding error map from different methods. Examples from DE X-ray soft-tissue generation (Left), Sparse-view CT reconstruction (Middle), and MRI T1-to-T2 synthesis are shown. The image quality metrics of each sample are indicated at the bottom left of the images.}
\label{fig:compare_method}
\end{figure*}

\vspace{0.1cm}
\noindent\textbf{Inference:} \textcolor{black}{Once the prior image generation network $f_{prior}(\cdot)$ of cGAN and the conditional diffusion network $f_{dm}(\cdot)$ are converged from training}, we can use them in CMDM for image translation. The overall inference pipeline of CMDM is illustrated in Figure \ref{fig:network}. Instead of starting the reverse process from a standard normal distribution $\mathcal{N}(y_T | 0, I)$ at $T$, the reverse process starts at a pre-defined time point $t_s \in [0, T]$ with 
\begin{equation} \label{eq:prior}
    \hat{y}_{t_s} = \sqrt{\gamma_{t_{s}}} y_{prior} + \sqrt{1-\gamma_{t_{s}}}\epsilon_{prior}
\end{equation}
where $y_{prior} = f_{prior} (x)$ and $\epsilon_{prior} \sim \mathcal{N}(0, I)$. $t_s$ is empirically set to $250$, depending on the translation application. By rearranging equation \ref{eq:addnoise}, we can approximate the target image $y_0$ as
\begin{equation}
    y_0 = \frac{y_t - \sqrt{1-\gamma_t} f_{dm} (x,  \hat{y}_t, \gamma_t) }{\sqrt{\gamma_t}}.
\end{equation}
Then, by substituting this estimation of $y_0$ into the posterior distribution of $q(y_{t-1} | (y_0, y_t))$ in equation \ref{eq:posterior}, each iteration of the reverse process can be formulated as
\begin{equation}
    y_{t-1} = \frac{1}{\sqrt{\alpha_t}} (y_t - \frac{1-\alpha_t}{\sqrt{1-\gamma_t}} f_{dm} (x, y_t, \gamma_t)) + \sqrt{1-\alpha_t} \epsilon_t
\end{equation}
where $\epsilon_t \sim \mathcal{N}(0, I)$. By starting the reverse process at shortcut time point $t=t_s$ with guidance from the prior image, the conditional diffusion model is closer to the endpoint, i.e. $t=0$, thus providing less diverged prediction from multiple predictions. To further improve the robustness, instead of only performing a single shortcut reverse path, we perform multiple shortcut reverse paths at $t_s$ with different noise initialization of $\epsilon_{prior}$ in equation \ref{eq:prior}, and ensemble these multi-path predictions by averaging
\begin{equation}
    y^{avg}_0 = \frac{1}{N_p} \sum^{N_p}_{p=1} y^p_0
\end{equation}
where $y^p_0$ is the prediction from a single shortcut reverse path and $N_p$ is the number of shortcut paths. To further refine the translation prediction, we perform this operation in a cascade style. To avoid over-fitting in the reverse process, we designed a residual averaging strategy for new prior image generation in the next cascade. Specifically, the new prior image is the averaged image from the previous prior image and the translated image from the last cascade. The full algorithm is summarized in Algorithm \ref{alg:test}. 

\begin{table*} [htb!]
\footnotesize
\centering
\caption{\textcolor{black}{Quantitative comparisons of translation results from different methods. I2I applications include DE X-ray image generation (soft-tissue and bone image), Sparse-view CT reconstruction (1/6 projection under-sampling and 1/4 projection under-sampling), and MRI inter-modality synthesis (T1-to-T2 and T1-to-FLAIR). The best results are marked in \textbf{bold}. "$\dagger$" means the differences between CMDM and all the previous baseline methods are significant at $p<0.002$. The averaged inference time of each method is reported in the right column.}}
\label{tab:compare_method}
    \begin{tabular}{l|c|c|c||c|c|c|c}
        \hline
        \textbf{DE X-ray}     & \multicolumn{3}{c}{\textbf{Soft-Tissue}}                   &  \multicolumn{3}{c}{\textbf{Bone}}               & \textbf{Average}     \Tstrut\Bstrut\\
        \cline{2-7}
        \textbf{Evaluation}   & PSNR             & SSIM              & MAE              & PSNR            & SSIM              & MAE           & \textbf{Time (Sec)}     \Tstrut\Bstrut\\
        \hline
        \hline
        UNet                  & $39.76\pm2.36$   & $0.984\pm0.003$     & $0.606\pm0.071$     & $41.33\pm3.18$   & $0.988\pm0.003$  & $0.571\pm0.066$  & 0.013     \Tstrut\Bstrut\\
        \hline
        cGAN                  & $39.82\pm2.37$   & $0.985\pm0.003$     & $0.603\pm0.072$     & $41.36\pm3.17$   & $0.988\pm0.003$  & $0.572\pm0.065$  & 0.013     \Tstrut\Bstrut\\
        \hline
        Palette v1            & $42.89\pm2.34$   & $0.987\pm0.002$     & $0.390\pm0.047$     & $43.06\pm3.16$   & $0.989\pm0.002$  & $0.373\pm0.042$  & 13.670     \Tstrut\Bstrut\\
        \hline
        Palette v2            & $43.11\pm2.34$   & $0.988\pm0.002$     & $0.382\pm0.045$     & $43.47\pm3.13$   & $0.990\pm0.002$  & $0.363\pm0.043$  & 273.420     \Tstrut\Bstrut\\
        \hline
        I2SB                  & $43.18\pm2.35$   & $0.988\pm0.002$     & $0.381\pm0.045$     & $43.49\pm3.14$   & $0.990\pm0.002$  & $0.367\pm0.043$  & 14.551     \Tstrut\Bstrut\\
        \hline
        BBDM                  & $43.08\pm2.35$   & $0.988\pm0.002$     & $0.382\pm0.044$     & $43.52\pm3.13$   & $0.989\pm0.002$  & $0.359\pm0.043$  & 15.121     \Tstrut\Bstrut\\
        \hline
        Ours                  & $\mathbf{44.27\pm2.33}$$^\dagger$   & $\mathbf{0.991\pm0.002}$$^\dagger$    & $\mathbf{0.369\pm0.041}$$^\dagger$    & $\mathbf{44.58\pm3.16}$$^\dagger$   & $\mathbf{0.992\pm0.003}$$^\dagger$    & $\mathbf{0.348\pm0.038}$$^\dagger$   & 154.663    \Tstrut\Bstrut\\
        \hline
    \end{tabular}

    \begin{tabular}{l|c|c|c||c|c|c|c}
        \hline
        \textbf{CT}           & \multicolumn{3}{c}{\textbf{1/6 Sparse-view}}                &  \multicolumn{3}{c}{\textbf{1/4 Sparse-view}}      & \textbf{Average}              \Tstrut\Bstrut\\
        \cline{2-7}
        \textbf{Evaluation}   & PSNR             & SSIM              & MAE              & PSNR            & SSIM              & MAE                 & \textbf{Time (Sec)}         \Tstrut\Bstrut\\
        \hline
        \hline
        UNet                  & $44.11\pm1.38$   & $0.977\pm0.004$     & $0.372\pm0.047$     & $46.32\pm1.27$   & $0.981\pm0.004$  & $0.315\pm0.040$   & 0.006       \Tstrut\Bstrut\\
        \hline
        cGAN                  & $44.13\pm1.39$   & $0.978\pm0.004$     & $0.370\pm0.047$     & $46.35\pm1.28$   & $0.981\pm0.004$  & $0.314\pm0.040$   & 0.006        \Tstrut\Bstrut\\
        \hline
        Palette v1            & $44.96\pm1.24$   & $0.980\pm0.003$     & $0.321\pm0.041$     & $46.75\pm1.26$   & $0.987\pm0.004$  & $0.310\pm0.039$   & 8.863       \Tstrut\Bstrut\\
        \hline
        Palette v2            & $45.56\pm1.24$   & $0.981\pm0.003$     & $0.318\pm0.040$     & $46.95\pm1.25$   & $0.988\pm0.003$  & $0.308\pm0.038$   & 177.202       \Tstrut\Bstrut\\
        \hline
        I2SB                  & $45.86\pm1.26$   & $0.982\pm0.003$     & $0.317\pm0.039$     & $46.91\pm1.26$   & $0.989\pm0.003$  & $0.309\pm0.039$   & 9.561       \Tstrut\Bstrut\\
        \hline
        BBDM                  & $45.73\pm1.24$   & $0.981\pm0.003$     & $0.318\pm0.040$     & $46.96\pm1.26$   & $0.989\pm0.003$  & $0.309\pm0.038$   & 9.987       \Tstrut\Bstrut\\
        \hline
        Ours                  & $\mathbf{46.42\pm1.22}$$^\dagger$   & $\mathbf{0.986\pm0.003}$$^\dagger$    & $\mathbf{0.302\pm0.039}$$^\dagger$    & $\mathbf{47.02\pm1.25}$$^\dagger$   & $\mathbf{0.990\pm0.003}$$^\dagger$    & $\mathbf{0.299\pm0.038}$$^\dagger$  &  108.821      \Tstrut\Bstrut\\
        \hline
    \end{tabular}
    
    \begin{tabular}{l|c|c|c||c|c|c|c}
        \hline
        \textbf{MRI}          & \multicolumn{3}{c}{\textbf{T1 $\rightarrow$ T2}}            &  \multicolumn{3}{c}{\textbf{T1 $\rightarrow$ FLAIR}}   & \textbf{Average}                 \Tstrut\Bstrut\\
        \cline{2-7}
        \textbf{Evaluation}   & PSNR             & SSIM              & MAE              & PSNR            & SSIM              & MAE                  & \textbf{Time (Sec)}      \Tstrut\Bstrut\\
        \hline
        \hline
        UNet                  & $27.17\pm1.56$   & $0.885\pm0.042$     & $0.222\pm0.051$     & $27.38\pm1.59$   & $0.891\pm0.046$  & $0.216\pm0.052$   & 0.006    \Tstrut\Bstrut\\
        \hline
        cGAN                  & $27.19\pm1.58$   & $0.887\pm0.044$     & $0.220\pm0.052$     & $27.41\pm1.58$   & $0.891\pm0.047$  & $0.217\pm0.053$   & 0.006    \Tstrut\Bstrut\\
        \hline
        Palette v1            & $27.52\pm1.57$   & $0.890\pm0.044$     & $0.218\pm0.051$     & $27.68\pm1.54$   & $0.897\pm0.046$  & $0.210\pm0.052$   & 8.863    \Tstrut\Bstrut\\
        \hline
        Palette v2            & $27.68\pm1.55$   & $0.891\pm0.043$     & $0.211\pm0.051$     & $27.79\pm1.52$   & $0.899\pm0.044$  & $0.206\pm0.051$   & 177.202    \Tstrut\Bstrut\\
        \hline
        I2SB                  & $27.85\pm1.56$   & $0.892\pm0.043$     & $0.208\pm0.051$     & $27.89\pm1.54$   & $0.898\pm0.043$  & $0.208\pm0.052$   & 9.561    \Tstrut\Bstrut\\
        \hline
        BBDM                  & $27.88\pm1.56$   & $0.892\pm0.043$     & $0.207\pm0.051$     & $27.86\pm1.53$   & $0.899\pm0.045$  & $0.207\pm0.051$   & 9.987    \Tstrut\Bstrut\\
        \hline
        Ours                  & $\mathbf{27.93\pm1.54}$$^\dagger$   & $\mathbf{0.898\pm0.042}$$^\dagger$    & $\mathbf{0.202\pm0.051}$$^\dagger$    & $\mathbf{27.98\pm1.54}$$^\dagger$   & $\mathbf{0.901\pm0.044}$$^\dagger$    & $\mathbf{.201\pm.051}$$^\dagger$  &  108.821    \Tstrut\Bstrut\\
        \hline
    \end{tabular}

\end{table*}

\subsection{Dataset Preparation}
We collected three medical image datasets with different medical image translation applications to validate our method. The first application is the image translation of conventional single-exposure chest radiography images into two-shot-based dual-energy (DE) images, which aims to reduce the expensive system cost of the DE system and higher radiation dose of two X-ray shots. Specifically, we collected 210 posterior-anterior DE chest radiographs with a two-shot DE digital radiography system \citep{zhou2019generation,wen2018enhanced}. The data was acquired using a 60 kVp exposure followed by a 120 kVp exposure procedure with 100 ms between exposures. The size of the images is $1024 \times 1024$ pixels. Based on this dataset, we further divide this task into two sub-tasks, including the translation of standard chest radiography into the soft-tissue image, and the translation of standard chest radiography into the bone image. The second application is image translation across MRI modalities, which aims to speed up the MRI acquisition that requires multiple protocols \citep{zhou2020dudornet}. Specifically, we collected an in-house MRI dataset consisting of 20 patients. We scanned each patient using three protocols, including T1, T2, and FLAIR, resulting in three 3D volumes of $320 \times 230 \times 18$ for each patient, and resized to $256 \times 256 \times 18$. $360$ 2D axial images are generated for each protocol. We further sub-divided this task into two components: translating the T1 image into the T2 image, and translating  the T1 image into the FLAIR image. The third application is the image translation of sparse-view CT (SVCT) images into full-view CT images, which aims to reduce the radiation dose in CT acquisition \citep{zhou2021limited,zhou2022dudodr}. We collected 10 whole-body CT scans from the AAPM Low-Dose CT Grand Challenge \citep{mccollough2016tu}. Each 3D scan contains $318 \sim 856$ axial slices covering a wide range of anatomical regions from chest to abdomen to pelvis, resulting in a total of $3397$ axial 2D images. Using the CT projection simulator, the fully sampled sinogram data was generated via $360$ projection views uniformly spaced between $0$ and $360$ degrees. Then, we uniformly sampled $90$ and $60$ projection views from the $360$ projection views, mimicking 4- and 6-fold projection view/radiation dose reduction. The paired full-view and sparse-view CT images were then reconstructed using Filtered Back Projection (FBP) based on these sinograms with the size of $256 \times 256$. For all three applications/datasets, we performed 5-fold cross-validation for evaluation considering their moderate scale. 

\begin{figure*}[htb!]
\centering
\includegraphics[width=0.86\textwidth]{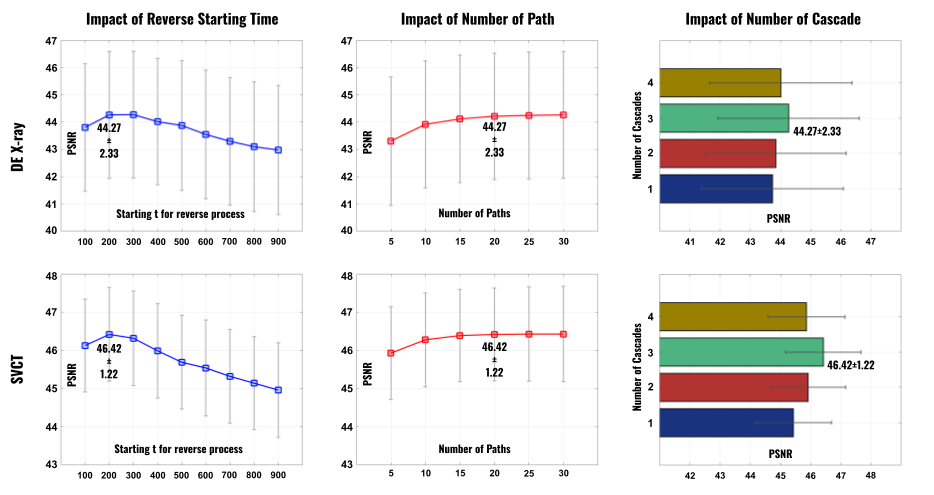}
\caption{Ablative studies on the reverse starting time (Left), the number of paths (Middle), and the number of cascades (Right). DE X-ray soft-tissue image generation and 1/6 SVCT reconstruction were utilized for these studies. Peak performances were annotated on the plots with the corresponding image quality metric, i.e. PSNR.}
\label{fig:abla}
\end{figure*}

\subsection{Evaluation Metrics and Baselines Comparisons}
To evaluate the translated image quality for the above-mentioned applications, we used the Peak Signal-to-Noise Ratio (PSNR), Structural Similarity Index (SSIM), and Mean Absolute Error (MAE) that was computed against their corresponding paired ground truth. For baseline comparisons, we compared our method's results against previous one-step CNN-based and diffusion-based image-to-image translation methods, including cGAN \citep{isola2017image}, Palette \citep{saharia2022palette}, Schrodinger Bridge Diffusion Model (I2SB) \citep{liu20232}, and Brownian Bridge Diffusion Models (BBDM) \citep{li2023bbdm}. Given that Palette utilizes random Gaussian noise as the initial input, we also compared two versions of Palette, including Palette with 1-sampling run (Palette v1) and Palette with 20-sampling runs with results averaging (Palette v2). I2SB and BBDM only have the 1-sampling run version given that there is no randomized input during sampling. Furthermore, we also conducted ablative studies on the hyper-parameters of CMDM, including the shortcut time point, number of shortcut paths, and number of cascades. 

\subsection{Implementation Details}
We implemented our method in PyTorch and performed experiments using an NVIDIA H100 GPU. We train all models with a batch size of 8 for 500k training steps. The Adam solver was used to optimize our models with $lr=1 \times 10^{-4}$, $\beta_1 = 0.9$, and $\beta_2 = 0.99$. We used an EMA rate of $0.9999$. A 10k linear learning rate warmup schedule was implemented. We used a linear noise schedule with 1000 time steps. 

\section{Experimental Results}
Figure \ref{fig:compare_method} shows qualitative comparisons between previous state-of-the-art and our methods. Examples from the DE X-ray dataset, SVCT reconstruction dataset, and MRI translation dataset are illustrated. For the DE X-ray example (left two columns), we can see all the previous translation methods can generate reasonable soft-tissue images, i.e. rib-suppression images, from the standard X-ray image. \textcolor{black}{While cGAN could generate visually plausible results with a PSNR of $44.74$dB}, the translated images still suffer from relatively inaccurate quantification as indicated by the error map. \textcolor{black}{On the other hand, we can see the previous diffusion-based methods, e.g. Palette and BBDM, both achieved significantly better translation as compared to cGAN with PSNR improving to $46.05$dB, with much fewer pixel-wise errors indicated by the error maps. In the last row, we can find that our CMDM further improved over the previous diffusion-based methods with PSNR reaching to $47.34$dB,} where further reduced pixel-wise error can be found in the cardiac and lung regions. Similarly, for the SVCT example (the middle two columns), cGAN can reasonably suppress the streak artifact in the input FBP SVCT image. However, significant residual errors can be found in the femoral head and pelvic bone regions. \textcolor{black}{On the other hand, we observe that the previous diffusion-based methods can suppress these errors, with PSNR reaching close to $46$dB. Furthermore, with CMDM combining cGAN and Diffusion, we can see that the overall error of our translation results are reduced even more, and the image quality is enhanced to PSNR of $46.23$dB.} Similar observations can be found for the T1-to-T2 translation example in the last two columns. 

The quantitative comparisons were summarized in Table \ref{tab:compare_method}. Similar to the observations from the visualizations, \textcolor{black}{we can see the traditional CNN-based approaches generally under-performed the diffusion-based approaches. For example, the cGAN only achieved an average PSNR of $39.82$dB and MAE of $0.603$ for the soft-tissue image translation, while the single reverse path Palette, i.e. Palette v1, significantly outperformed it with PSNR of $42.89$dB and MAE of $0.390$. Running multiple reverse paths of Palette and averaging the outputs, i.e. Palette v2, led to improved performance which reached similar performances of I2SB and BBDM with PSNR of $43.11$dB and MAE of $0.382$. In the last row, our CMDM achieved an average PSNR of $44.27$dB and MAE of $0.369$} that significantly outperformed all the previous baseline methods. Comparing the soft-tissue image translation task to the bone image translation task, the CMDM had slightly higher performance on the latter task since the bone image without complex soft-tissue texture can be relatively easier to generate as compared to the soft-tissue image. For the inference speed in DE X-ray applications, I2SB and BBDM with a single reverse path took an average of $14.55$ and $15.12$ seconds, respectively. CMDM with the best performance took an average of $154.66$ seconds per inference since multiple shortcut reverse paths are needed. Similar to the quantitative results for the DE X-ray, we found our CMDM consistently outperformed previous CNN and diffusion-based baseline methods for both the SVCT reconstruction applications and the MRI inter-modality translation applications. 

\begin{figure*}[htb!]
\centering
\includegraphics[width=0.975\textwidth]{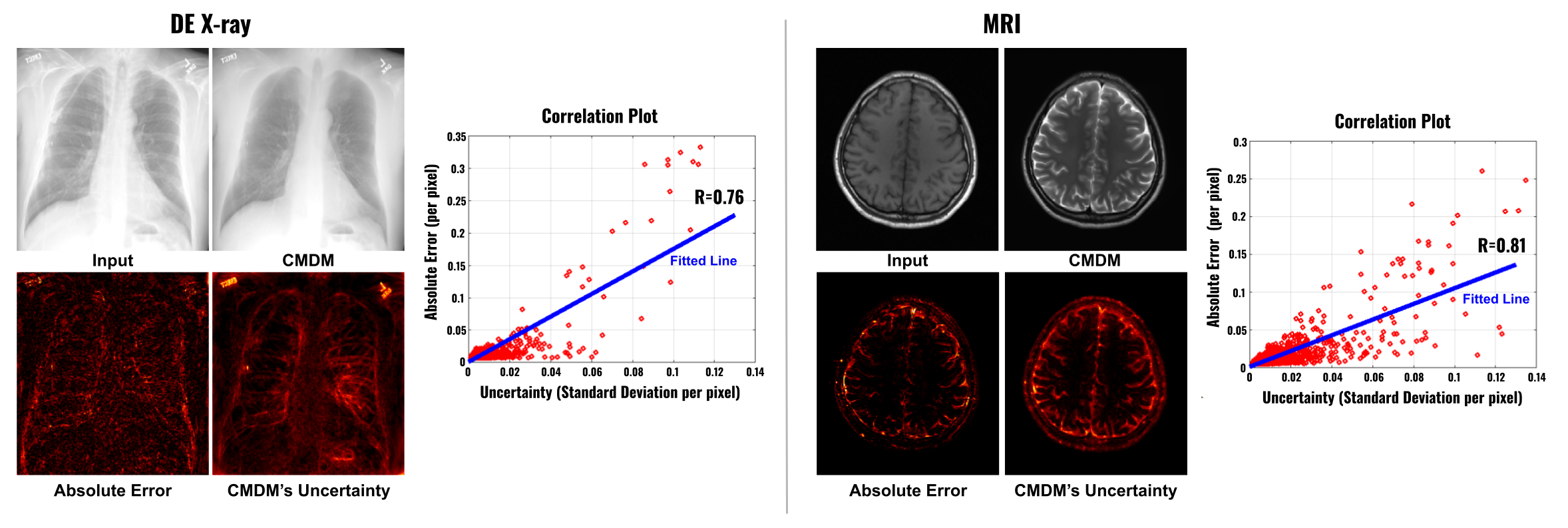}
\caption{Examples of CMDM's uncertainty estimation for DE X-ray soft-tissue image generation (Left) and MRI T1-to-T2 synthesis (Right). The relationship plots between the absolute error (bottom left) and the uncertainty (bottom right) were shown as well. Positive correlations with $R>0.75$ were found for both cases.}
\label{fig:uncertainty}
\end{figure*}

\begin{table} [htb!]
\footnotesize
\centering
\caption{\textcolor{black}{Quantitative comparison of CMDM with different prior strategies. Analysis with DE X-ray soft-tissue generation task, 1/6 SVCT reconstruction task, and T1-to-T2 MRI synthesis task are reported.}}
\label{tab:prior}
    \begin{tabular}{l|c|c|c}
        \hline
        \textbf{MAE}          & DE X-Ray             & SVCT              & MRI              \Tstrut\Bstrut\\
        \hline
        \hline
        w/o prior             & $0.379\pm0.043$   & $0.316\pm0.042$     & $0.210\pm0.051$     \Tstrut\Bstrut\\
        \hline
        UNet prior            & $0.370\pm0.041$   & $0.306\pm0.041$     & $0.203\pm0.051$      \Tstrut\Bstrut\\
        \hline
        UFNet prior           & $0.366\pm0.041$   & $0.303\pm0.040$     & $0.201\pm0.050$      \Tstrut\Bstrut\\
        \hline
        cGAN prior            & $0.369\pm0.041$   & $0.302\pm0.039$     & $0.201\pm0.051$      \Tstrut\Bstrut\\
        \hline
    \end{tabular}
\end{table}

We conducted ablative studies for the hyper-parameters in CMDM, including the reverse starting time, the number of shortcut paths, and the number of cascades. The results for the DE X-ray and SVCT are summarized in Figure \ref{fig:abla}. First, for the reverse starting time, we can see that setting $t_s$ to around $200$ yields the best performance, and the performance starts to degrade if we further increase it. It is worth noticing that using $t_s=200$ here not only yields the best performance but allows us to reduce the inference time by about 5 times as compared to the previous diffusion methods that start at $t=1000$ or beyond. Second, for the number of shortcut paths, we can see that the performance increases as we use an increasing number of paths. The performance started to converge when $20$ paths were used. Because the inference time increased linearly as we increased the number of paths, we chose the converging point $N_p=20$ in our method. Thirdly, for the number of cascades, we found that the performance gradually boosted as the number of cascades increased. However, peak performance was reached when $N_c=3$, and the inference started to overfit, leading to degraded translation performance. \textcolor{black}{Lastly, we investigated the impact on CMDM when different prior image generations were used, including priors from UNet \citep{ronneberger2015u}, Under-to-fully-complete Network (UFNet \citep{zhou2022dudoufnet}), and cGAN \citep{isola2017image}. As we can see from Table \ref{tab:prior}, using CMDM with prior always outperforms CMDM without prior. Among all the prior generated, CMDM with priors generated from cGAN and UFNet yields the best performance. Moreover, we also studied CMDM with or without the conditional input for the diffusion part. As we can see from Table \ref{tab:cond}, CMDM without conditional input can still generate a reasonable translation guided by the prior image. However, CMDM with condition input with more translation guidance still yields the best performance. }

\begin{table} [htb!]
\footnotesize
\centering
\caption{\textcolor{black}{Quantitative comparison of CMDM with or without images to be translated as conditional inputs in the diffusion part. Analysis with DE X-ray soft-tissue generation task, 1/6 SVCT reconstruction task, and T1-to-T2 MRI synthesis task are reported.}}
\label{tab:cond}
    \begin{tabular}{l|c|c|c}
        \hline
        \textbf{MAE}          & DE X-Ray             & SVCT              & MRI              \Tstrut\Bstrut\\
        \hline
        \hline
        w/o condition         & $0.517\pm0.059$   & $0.339\pm0.043$     & $0.219\pm0.053$     \Tstrut\Bstrut\\
        \hline
        w condition           & $0.369\pm0.041$   & $0.302\pm0.039$     & $0.202\pm0.051$      \Tstrut\Bstrut\\
        \hline
    \end{tabular}
\end{table}

In addition to the translation performance, we also conducted ablative studies on CMDM's uncertainty estimation. Two examples of DE X-ray and MRI T1-to-T2 translation are shown in Figure \ref{fig:uncertainty}. On the bottom, both the pixel-wise absolute error and the pixel-wise uncertainty (i.e. computed by the standard deviation of multiple shortcut path predictions) are visualized. The corresponding scatter plot of their pixel-wise relationship is also shown on the right. We found that the pixel-wise uncertainty and the absolute error have a good correlation. For the DE X-ray example and the MRI example here, we have a correlation coefficient equal to $0.76$ and $0.81$, respectively. This is particularly useful when ground truth is unavailable to compute the translation error, where uncertainty can indicate the potential error distributions. The correlation of the pixel-wise uncertainty and the absolute error for the whole test set is summarized in Table \ref{tab:compare_uncertainty}. By running multiple sampling runs of Palette \citep{saharia2022palette}, i.e. Palette v2, it can also produce the pixel-wise standard deviation for uncertainty estimation. In Table \ref{tab:compare_uncertainty}, we can see CMDM achieving a better-averaged correlation across all three translation applications. 

\begin{table} [htb!]
\footnotesize
\centering
\caption{Averaged correlation of the pixel-wise absolute error and the pixel-wise uncertainty, i.e. computed by the standard deviation of multiple paths' predictions. DE X-ray soft-tissue generation task, 1/6 SVCT reconstruction task, and T1-to-T2 MRI synthesis task are reported.}
\label{tab:compare_uncertainty}
    \begin{tabular}{l|c|c|c}
        \hline
        \textbf{Correlation}   & DE X-Ray             & SVCT              & MRI              \Tstrut\Bstrut\\
        \hline
        \hline
        Palette v2            & $0.678\pm0.162$   & $0.702\pm0.137$     & $0.676\pm0.108$     \Tstrut\Bstrut\\
        \hline
        CMDM                  & $0.695\pm0.142$   & $0.718\pm0.127$     & $0.687\pm0.089$      \Tstrut\Bstrut\\
        \hline
    \end{tabular}
\end{table}

\section{Discussion} 
In this work, we developed a novel image translation method, called CMDM, that efficiently integrates GAN and DM to enable high-quality medical image-to-image translation. There are several key advantages of this method. First, we utilized a previous CNN-based translation method to generate a virtual "$t=0$" image for the diffusion model. This image is added with the scheduled noise, so we can start the diffusion reverse process at a scheduled shortcut time point. As illustrated in Figure \ref{fig:previous}, initializing the reverse process with pure noise may lead to sub-optimal results, while here, starting the reverse process with a roughly estimated image (e.g. cGAN's prediction) with the scheduled noise not only can help stabilize the reverse sampling process, but also reduce the required number of reverse iterations, i.e shorten the inference time. Second, instead of adding one noise schedule \citep{chung2022come,gao2023corediff}, we added different noises to this "$t=0$" image and performed the same reverse process multiple times in each cascade. The corresponding cascade output is simply the averaged outputs from these paths. This averaging operation inherently reduces the randomness from the different noises and thus improves the translation robustness. Based on results from multiple reverse runs, we can generate pixel-wise uncertainty estimation for the translation results, which is also a key advantage. Lastly, we also devised a cascade framework with a residual averaging strategy. This design helps us enhance performance without training additional models, but may come at the cost of additional inference time. It is worth noticing that our CMDM can be viewed as a plug-and-play module that helps improve the performance of cGAN, i.e. the one-step inference model used in CMDM, as shown in Table \ref{tab:compare_method}. Ideally, our approach can also be added as a plug-and-play module to the other previous translation methods for potential translation performance improvements. 

We collected three medical image datasets with a total of six different medical image translation tasks to validate our method. From our experimental results, we demonstrated our method can generate high-quality translated images that consistently outperformed previous baseline methods (Figure \ref{fig:compare_method} and Table \ref{tab:compare_method}). For example, we can see that CMDM achieved PSNR $>44$dB for both DE soft-tissue image generation and DE bone image generation, while all the previous methods are below $44$dB. Although CMDM achieves the best performance, it requires a relatively longer inference time as compared to previous methods that need a single reverse run. For example, CMDM needs $154.66$ seconds on average for the DE X-ray application, but Palette v1, I2SB, and BBDM only need about $13$ seconds. \textcolor{black}{However, we can either reduce the number of cascades or the number of shortcut paths in CMDM to balance the computation time and performance need. The default settings in our CMDM are $N_c=3$ and $N_p=20$. According to the studies reported in Figure \ref{fig:abla}, we could reduce the number of cascades ($N_c$) to $1$ to shorten the inference time by nearly three times which would result in PSNR=$43.75$dB. This result still outperformed all the previous baseline methods (Table \ref{tab:compare_method}). Similarly, we could also reduce the number of shortcut paths ($N_p$) to $10$ to cut the inference time by nearly half and still outperform all the previous baseline methods. On the other hand, we believe these hyper-parameters also need to be tuned for different translation applications to find the optimal balance between performance and computation/time budgets.} Besides the translation itself, CMDM also generates pixel-wise uncertainty estimation. As we can see from Figure \ref{fig:uncertainty} and Table \ref{tab:compare_uncertainty}, CMDM's uncertainty estimation demonstrated good correlations with the absolute error that can only be computed when the ground truth is available. Since ground truth is commonly unavailable when deployed in clinical scenarios for estimating the error, we believe our uncertainty estimation is potentially useful for the user to decide which region is trustworthy for downstream applications, such as diagnosis and treatment planning.

The presented work also has limitations with several potential improvements that are important subjects of our future studies. Firstly, we only validated our method on three different modalities, and evaluations on more diverse applications could be included. Even though we framed CMDM as an image-space post-processing tool here, we believe it can be further tailored to specific translation problems. For example, we could include physic-informed modules, such as data consistency \citep{schlemper2017deep,song2021solving}, in CMDM which may further improve its applications in medical image reconstruction \citep{zbontar2018fastmri,sidky2022report}. \textcolor{black}{Secondly, the current CMDM is implemented in a 2D fashion, while 3D is desirable in many medical image translation tasks. Theoretically, we could directly change all the networks in CMDM into 3D networks to enable 3D applications, but it may be infeasible with the current computation resources. For example, we attempted to employ the 3D CMDM with an input size of $256\times256\times128$ on an 80G H100 GPU, however, it cannot fit into the memory even with a single batch size. Alternatively, we could also utilize multi-view diffusion or 2.5D or memory-efficient strategies to scale CMDM into 3D \citep{chung2023solving,xie2023dose,bieder2024denoising,chen20242} which will be extensively investigated in our future works. Thirdly, the inference speed is still considered relatively long as compared to previous methods, especially the classic CNN-based methods. While we discussed the trade-off between performance and speed in the previous paragraph, it is also desirable to maintain optimal performance with increased inference speed. Utilizing accelerated diffusion models, such as DDIM and Resshift \citep{song2020denoising,yue2024resshift}, in CMDM could potentially help achieve this goal. To accelerate the inference speed for time-critical clinical scenarios, such as real-time translation in intervention radiology, one could also consider alternative solutions. For example, we could also consider distilling the diffusion model knowledge into the one-step inference GAN model \citep{kang2024distilling}, such that GAN with diffusion model performance and real-time capability can be realized. Fourthly, in the current implementation of CMDM, we did not implement ways to monitor the first step’s image generation. If unsatisfactory results were generated in the first step, the error could propagate to the next step. However, this should be reflected on the CMDM final uncertainty map where increased uncertainty value, i.e. pixel-wise standard deviation, should be observed. On the other hand, we could also further include uncertainty estimation techniques, e.g. Monte Carlo Dropout \citep{gal2016dropout}, in the first step of cGAN, thus monitoring prior image generation. Lastly, CMDM requires data with paired images for training, but such data may not always be available in certain applications. Unpaired translation diffusion model strategies \citep{sasaki2021unit,ozbey2023unsupervised} could also potentially be deployed here to mitigate this challenge. For example, one could consider using CycleGAN to generate the prior image, and then using a multi-path version of UNIT-DDPM \citep{sasaki2021unit} for further refinement of the prior image. This is an interesting direction to be investigated in our future works. Moreover, future works also include evaluations of how CMDM impacts the downstream clinical applications. For example, we will investigate if the CMDM-translated images provide similar lesion detection capability or radiomic features when compared to the ground truth images, thus validating the clinical values of our method.}

\section{Conclusion} 
Our work proposes a Cascaded Multi-path Shortcut Diffusion Model (CMDM) - a simple and novel strategy for high-quality medical image-to-image translation. The proposed method first utilizes a classic CNN-based translation method to generate a prior image. By adding different noises to this image, we then run multiple reverse samplings starting with the noisy images, i.e. shortcuts. With this process in each cascade, the translation output is obtained by averaging them, and the uncertainty estimation is obtained by calculating the standard deviation. Based on this, a cascade framework with residual averaging is further proposed to gradually refine the translation. For validation, we utilized three medical image datasets across X-ray, CT, and MRI. Our experimental results showed that CMDM can provide high-quality translation results, better than previous translation baselines for different sub-tasks. In parallel, CMDM also provides reasonable uncertainty estimations that correlate well with the translation error maps. We believe CMDM could be potentially adapted to other applications where both high-quality translation and uncertainty estimation are required. 

\section*{Acknowledgments}
This work was supported by the National Institutes of Health (NIH) grant R01EB025468 and grant R01CA275188.

\section*{Declaration of Competing Interest}
The authors declare that they have no known competing financial interests or personal relationships that could have appeared to influence the work reported in this paper.

\section*{Credit authorship contribution statement }
\textbf{Yinchi Zhou}: Conceptualization, Methodology, Software, Visualization, Validation, Formal analysis, Writing original draft.
\textbf{Tianqi Chen}: Results analysis, Writing - review and editing.
\textbf{Jun Hou}: Results analysis, Writing - review and editing.
\textbf{Huidong Xie}: Conceptualization, Methodology, Software, Writing - review and editing.
\textbf{Nicha C. Dvornek}: Writing - review and editing
\textbf{S. Kevin Zhou}: Data preparation, Writing - review and editing.
\textbf{David L. Wilson}: Data preparation, Writing - review and editing.
\textbf{James S. Duncan}: Writing - review and editing
\textbf{Chi Liu}: Writing - review and editing
\textbf{Bo Zhou}: Conceptualization, Methodology, Software, Visualization, Validation, Formal analysis, Writing original draft, Supervision.

\bibliographystyle{model2-names.bst}\biboptions{authoryear}
\bibliography{refs}

\end{document}